\documentclass[12pt,preprint]{aastex}
\usepackage{natbib}
\usepackage{graphics}
\usepackage{graphicx}
\usepackage{epsfig}
\usepackage{amssymb}

\begin{document}

\title{Giant Radio Pulses from the Crab Pulsar}

\author{A. Jessner}
\affil{Max Planck Institute for Radioastronomy, Auf dem H\"ugel 69,
D-53121, Bonn, Germany}
\email{jessner@mpifr-bonn.mpg.de}

\author{A. S\l{}owikowska}
\affil{Copernicus Astronomical Center,
        Rabia{\' n}ska 8, 87-100 Toru{\' n}, Poland}
\email{aga@ncac.torun.pl}

\author{B. Klein}
\affil{Max Planck Institute for Radioastronomy, Auf dem H\"ugel 69,
D-53121, Bonn, Germany}
\email{bklein@mpifr-bonn.mpg.de}

\author{H. Lesch}
\affil{Institut f\"ur Astronomie und Astrophysik LMU, 
Scheinerstr. 1,  D-81679 M\"unchen, Germany}
\email{lesch@usm.uni-muenchen.de}

\author{C.H. Jaroschek}
\affil{Institut f\"ur Astronomie und Astrophysik LMU, 
Scheinerstr. 1,  D-81679 M\"unchen, Germany}
\email{cjarosch@usm.uni-muenchen.de}

\author{G. Kanbach}
\affil{Max Planck Institute for Extraterrestrial Physics,
Postfach 1312, D-85741 Garching, Germany}
\email{gok@mpe.mpg.de}

\author{T.H. Hankins}
\affil{Physics Department, New Mexico Institute of Mining and 
Technology, Soccoro, NM 87801, USA}
\email{thankins@aoc.nrao.edu}

\begin{abstract}
Individual giant radio pulses (GRPs) from the Crab pulsar last only  a few
microseconds. However, during that time they rank among the brightest
objects in the radio sky reaching peak flux densities of up to 1500 Jy even
at high radio frequencies. Our observations show that GRPs can be found in
all phases of ordinary radio emission including the two high frequency components
(HFCs) visible only between 5 and 9~GHz \citep{moff96}. 
This leads us to believe that there is no difference in the emission mechanism
of the main pulse (MP), inter pulse (IP) and HFCs. 
High resolution dynamic spectra from our recent observations of
giant pulses with the Effelsberg telescope at a center 
frequency of 8.35~GHz show distinct spectral maxima within our observational
bandwidth of 500~MHz for individual pulses. Their narrow band
components appear to be brighter at  higher frequencies (8.6~GHz)
than at  lower ones (8.1~GHz). Moreover, there is an evidence
for spectral evolution within and between those structures.
High frequency features occur earlier than low frequency ones.
Strong plasma turbulence might be a feasible mechanism for the
creation of the high energy densities of 
$\sim 6.7 \times 10^4~\rm erg~cm^{-3}$
and brightness temperatures  of $\sim 10^{31}~\rm K$. 
\end{abstract}


\section{Observational Techniques}

\label{obs}

Observations with the Effelsberg 100-m radio telescope began on
25 November 2003 and ended 
on  28 November 2003. We used a  secondary focus cooled  HEMT receiver with a center
frequency of 8.35 GHz, providing two circularly polarized IF signals with a  system
temperature of  25~K on both channels.
With a sky temperature of 8~K, and a contribution of 33~K from the Crab nebula,
the effective system temperature was 70~K.
Two detection systems were used.
First, a polarimeter with 1.1~GHz bandwidth detected total power
of the left-hand 
and right-hand circularly polarized signals (LHC and RHC respectively) as well as 
$\cos \angle (\rm~LHC, RHC)$ and $\sin \angle (\rm~LHC, RHC)$.
These four signals were then recorded by the standard 
{\bf E}ffelsberg {\bf P}ulsar {\bf O}bservation {\bf S}ystem ({\bf EPOS}). 
With the Crab pulsar's dispersion measure of $56.8~{\rm pc~cm^{-3}}$, we had 
an un--dedispersed time resolution of $t_{sample} = 890~\mu{\rm s}$,
while our sampling resolution was fixed at $\sim 640~\rm \mu s$.
EPOS was therefore set to continuously record data blocks containing 20 periods
divided into 1020 phase bins for all four signals (Fig.~1).

The minimum detectable flux per bin was $\Delta S_{min} = 0.117~{\rm Jy}$. 
Considering the dispersion pulse broadening, the detection limit for a
single GRP of duration $\tau_{grp} = 3~\mu{\rm s}$
would be $\Delta S_{min}\,t_{sample}/\tau_{grp} = 25~{\rm Jy}$. Selected
GRPs are shown on Fig.~2.

A wide band ultra-high resolution detection system similar to the one of 
\citet{Han2003} was used as well.
The two IF signals (100--600~MHz) carrying the LHC and RHC
information were sampled and  recorded with a fast digital storage
oscilloscope (LeCroy LC584AL). 
$4 \times 10^{6}$ samples per channel were recorded at the rate of
$2 \times 10^{9}$ samples per second. 
The storage scope was used in a single shot mode. 
We used giant radio pulses to trigger data acquisition by the digital
storage scope. The RHC signal was detected, amplified and low-pass filtered (3~kHz),
then passed to a second scope (Tektronics 744). The second scope triggered the digital
scope whenever pulses stronger than 3--4 times rms (75--100~Jy) were detected.
The recorded waveforms were then 
transferred to disk. This required manual intervention and led to a dead time
of about 100 seconds after each pulse.  

\section{EPOS Data}
\label{epos}
About $2.4 \times 10^6$ periods were observed with EPOS. 
Presented results are only based on
a third of all observed rotations.
This selection was mostly caused by an insufficient quality
of the data.
Because of the Crab pulsar's weak signal at  8.35~GHz and the strong nebula background,
ordinary single pulses were not observable with Effelsberg at that frequency. 
It takes about 20 min integration to assemble a mean profile at that frequency.    
The giant pulses come in short outbursts, about 5 to 20 minutes in duration,
and appear extremely prominent during such burst phases.

In our subsequent analysis, we set a threshold level of 5 rms  =  125 Jy on
the sum of RHC and LHC for the same phase bin to count as a detection of 
a giant pulse. More than 1300 giant pulses were
detected that way  (Fig.~2) and their arrival phases were computed by using
the TEMPO{\footnote{http://pulsar.princeton.edu/tempo}}
pulsar timing package.
The data were aligned using the current Jodrell Bank timing model. They were 
found to be a perfect match to the time of arrivals (TOAs) obtained at Jodrell Bank
before and after the Effelsberg observing session. Upper
panel of Fig.~3 shows these phases. We found that the giant pulses
occur at all those phases where the radio components of the Crab
pulsar emission are observed (Fig.~3, bottom panel).

For all giant pulses, i.e. regardless of their phases, the histogram
of their peak strengths at 8.35 GHz can be described by a power law with a
slope $\sim -3.34 \pm 0.19$ (Fig. 4). This result is consistent
with the results obtained by others \citep[e.g.][]{Lundgren95}.
Moreover in Fig.~5 , we show the phase resolved distributions of four components:
two high frequency components (HFC1 and HFC2), and the main and
inter pulse components (MP and IP).
Because of limited statistics we could make a model fit only for the IP component,
where a power law index $\sim -3.13 \pm 0.22$ was found, 
and it is consistent with the value of $\sim -2.9$ presented by \cite{Cordes04}.
So far, there is no evidence that the distributions for HFC1 and HFC2
differ from each other. However, their slopes seems to be steeper
than the slope for the IP component.

The results obtained so far by us suggest that the physical
conditions in the regions responsible for HFC emission might be  similar to those
in the main pulse (MP) and inter pulse (IP) emission regions.
We plan to pursue this question further in the beginning of 2005, when a new
observation campaign at two frequencies, 4.85~GHz and 8.35~GHz,
of the GRPs of the Crab pulsar is scheduled.

\section{High Resolution Detections}
Because of the burst-like character of the giant pulse emission and the long dead-times in 
our data acquisition, only 150 pulses were observable with the high resolution equipment
(for example Fig.~6). The peak strength varied from our threshold of $\sim$~75~Jy
to a few rare events which even  exceeded 1000~Jy. We computed dynamic spectra
for the LHC and RHC signals by successive Fourier transforms and squaring of the data.
Baseline and sensitivity corrections were also applied  through the use of bandpass averages.  
The pulses were typically 2--5~$\mu s$ wide and certainly wider than the resolution
of our incoherent software de-dispersion, $0.7~\mu{\rm s}$. \citet{Han2003} showed the 
strong {\it temporal}  variabilities within the giant pulses,
which are possibly unresolved because of the limited  bandwidth of the observations.
Although giant pulses can be received over a wide bandwidth,
the individual pulses do not have a uniform spectrum, as seen in Fig.~7. The
strongest emission was predominantly detected in the upper quarter of the
accessible  
bandwidth. The individual pulses consist of $\sim~100$~MHz-wide clusters of narrow 
$\delta\nu~\sim~2$~MHz spectral 
lines waxing and waning with time (Fig.~7, upper).
High frequency features appear earlier than low frequency features.
Furthermore, if two giant pulses occur in rapid succession (Fig.~6,~7) 
separated by only $100~\mu{\rm s}$, their spectra are similar though 
not identical. The maximum emission of the leading pulse occurs at higher frequencies
than that of the trailing pulse (Fig.~6, bottom). The separation of spectral maxima also 
decreases in the trailing pulse. Intrinsic fluctuations of the emission
process or - alternatively - scintillation could be the cause.

\section{Characteristics of the Emission Mechanism}
\label{emission}
The maximum flux of $S_{\nu grp}$  =  $1500$~Jy is detected on a typical timescale 
of $\tau_{grp}$ = $3 \times 10^{-6}~{\rm s}$. With the Crab's distance of
$d = 2~\rm kpc$
and accounting for $50\,\%$ bandwidth coverage
($\epsilon_\nu$ = $0.5$, $\Delta\nu$ = $1.1~\rm GHz$) a peak luminosity of
$L_{grp} = S_{\nu grp} \epsilon_\nu 
 \Delta \nu d^2$ $[1-\cos(\tau_{grp}/P)] = 1.6 \times 10^{25}\rm~erg~s^{-1}$ is
obtained. The pulse duration limits the size of the
emitting volume to $V_{grp} \sim (c\tau_{grp})^3 = 7 \times 10^{15}~{\rm cm^3}$ 
and a rough estimate of the dissipated energy density yields 
$L_{grp} \tau_{grp} V_{grp}^{-1} \sim 7 \times 10^4 {\rm~erg~cm^{-3}}$.
For our observations the brightness temperature is estimated to 
$T_{b}$ = $4.4 \times 10^{31}{\rm~K}$
applying the relation $T_{b} = 2\pi S_{\nu grp}\tau_{grp} d^2 /(k \nu^2
\tau_{grp}^3)$ in accordance with \citet{Sog04}. Such extreme brightness temperatures
accompanied by strong polarization indicate
a highly coherent radio emission process. 
We find that the GRPs occur at the same phases as the ordinary radio emission,
perhaps dominating the radio profile at high frequencies. 
Such coincidence suggests that the GRPs are of the same origin as the radio
emission at $x \sim 20~r_{ns}$. At these emission heights, the interaction 
of relativistic particles with  
turbulent plasma wave packets in the pulsar magnetosphere represents a possible
source process for the creation of GRPs \citep{Han2003}. Strong plasma
turbulence is the highly non-linear dynamic equilibrium between density
fluctuations and radiative plasma modes. The purely electrostatic two-stream
instability parallel to the pulsar dipolar field serves as the trigger
mechanism. In the pulsar magnetosphere, electron-positron pairs are created in
events which are sharply localized in space and time. Each event creates an
individual plasma shell. 
Different shell generations have different bulk speeds and can
interpenetrate the magnetosphere on the way outwards \citep{Jaro2004}.
In such colliding plasma shells, the plasma two-stream
instability is excited, which evolves to a fully saturated equilibrium of
turbulent plasma modes.    
Self-induced, strong density fluctuations in the high-energy beam are coupled to 
density fluctuations in the low-energy background population via radiation 
pressure.
Such coupling leads to significant 
bunching of particles and thereby to coherent emission expected at a frequency of
$\gamma^2\nu_{pe}$, where $\gamma$ denotes the Lorentz factor in the frame of the slower
plasma shell and $\nu_{pe}$ is the plasma frequency of an {\it intrinsically}
relativistic plasma.

The mechanism outlined above is attributed as {\it collisionless
Bremsstrahlung} \citep{wea91} and is identified
in self-consistent 3D particle-in-cell (PIC)
simulations \citep{Jaro2004}. In these simulations, coherent radiation
emission stimulated by strong Langmuir turbulence is studied in slab geometry
for similar conditions (particle densities $\sim$ corotation densities of
$10^9~\rm cm^{-3}$  and 
Lorentz factors of $\gamma = 6$). Turbulent energy densities reach up
to $\sim 10^5~\rm erg~cm^{-3}$ with rise times and fluctuation timescales
of a few nanoseconds.  The emission of a fraction of this energy density is
sufficient for the observed GRPs. The  PIC simulations show that the emitted
radiation has a Poynting flux with magnitude, temporal structure and linear
polarization in good agreement with our observations. The radiation would last
until the supply of high energy particles is exhausted.

\section*{Acknowledgments} 

We would like to thank Michael Kramer for the kind supply of
ephemeris and TOAs from the Crab pulsar and H. Wiedenh\"over and K. Grypstra
for their help with the LeCroy Scope and analog signal processing. Aga
S\l{}owikowska was supported by grant KBN 2P03D.004.24. She 
would like to thank Bronek Rudak for his very bright suggestions.
We are pleased to acknowledge two anonymous referees for their
comments.




\newpage

\begin{figure}
\begin{center}
\includegraphics[angle =  270, scale =  0.74]{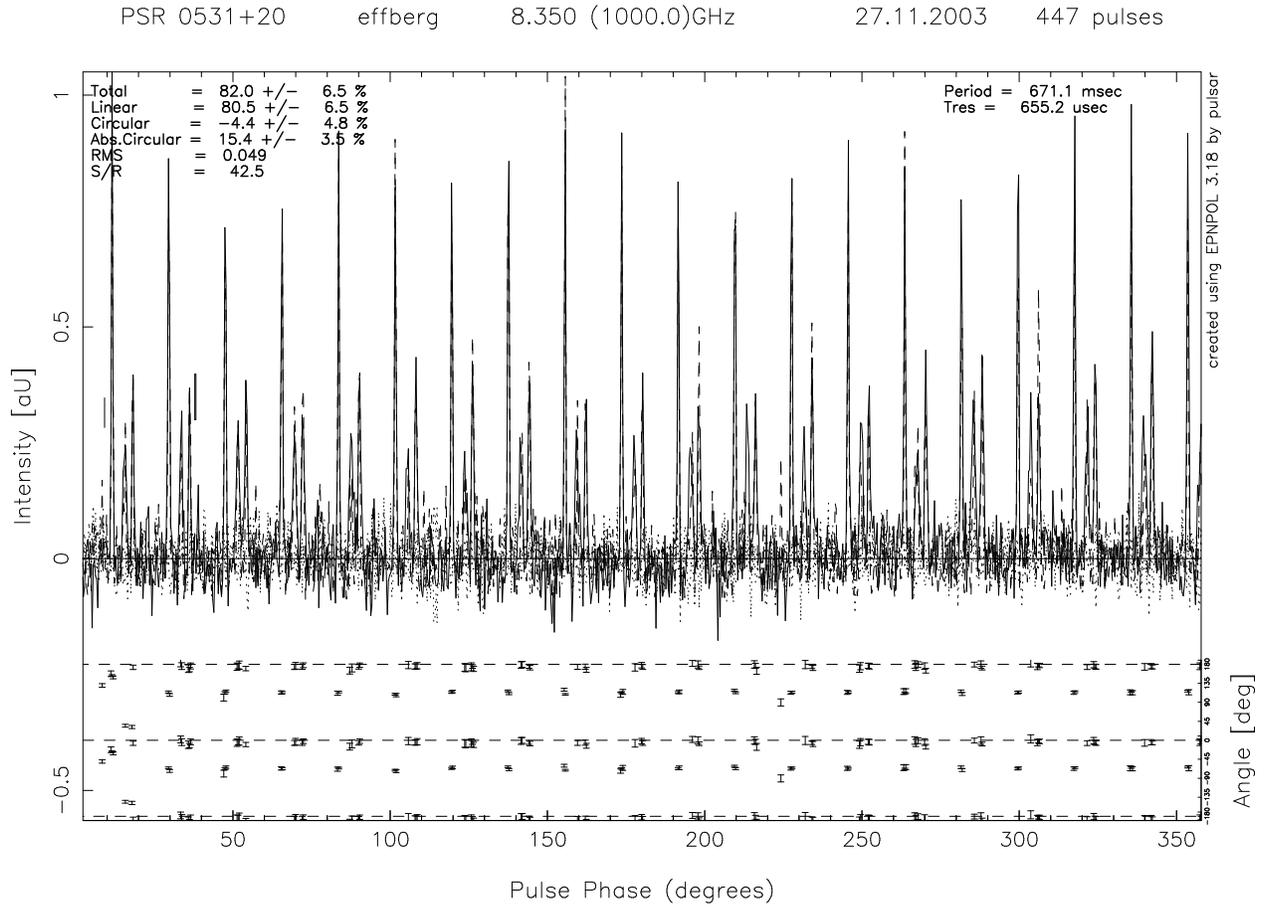}
\caption{Average pulse profiles of PSR~B0531+21 at 8.35~GHz. Twenty periods and
corresponding polarisation characteristics are shown.}
\end{center}
\end{figure}

\newpage

\begin{figure}
\begin{center}
\includegraphics[angle =  90, scale =  0.6]{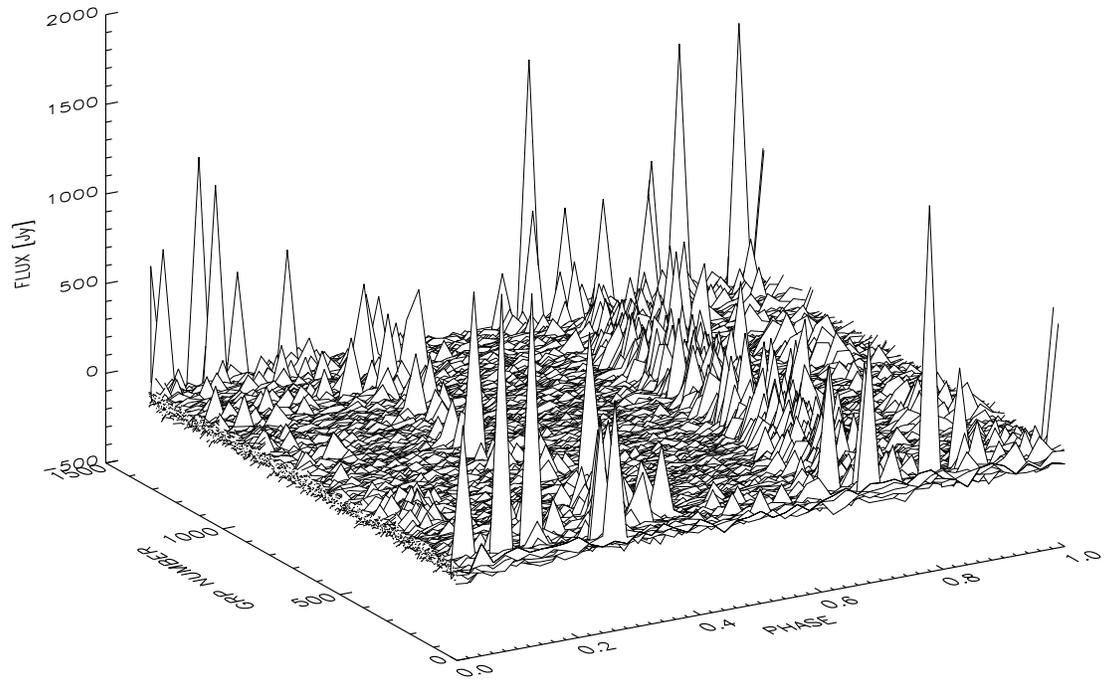}
\end{center}
\caption{Train of strong giant pulses at 8.35~GHz (EPOS, $\Delta\nu  =  1.1~{\rm
GHz}$, sum of the right and left handed circular polarisation signals).
Results obtained from  6.7 hr of observation with the 100~m Effelsberg
telescope.}
\end{figure} 

\newpage

\begin{figure}
\begin{center}
\includegraphics[angle = 180,scale =  0.46]{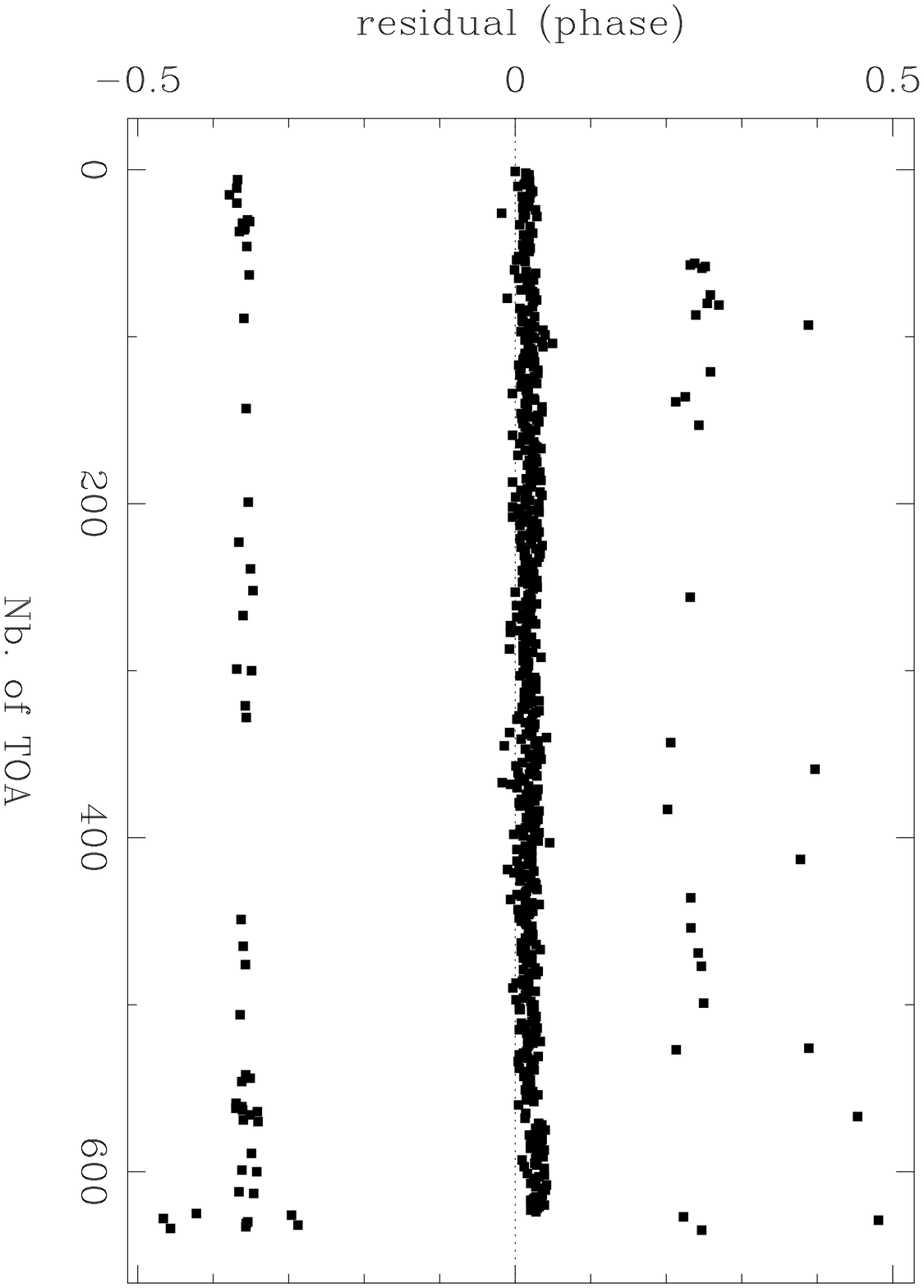}
\includegraphics[angle = 90,scale =  0.5]{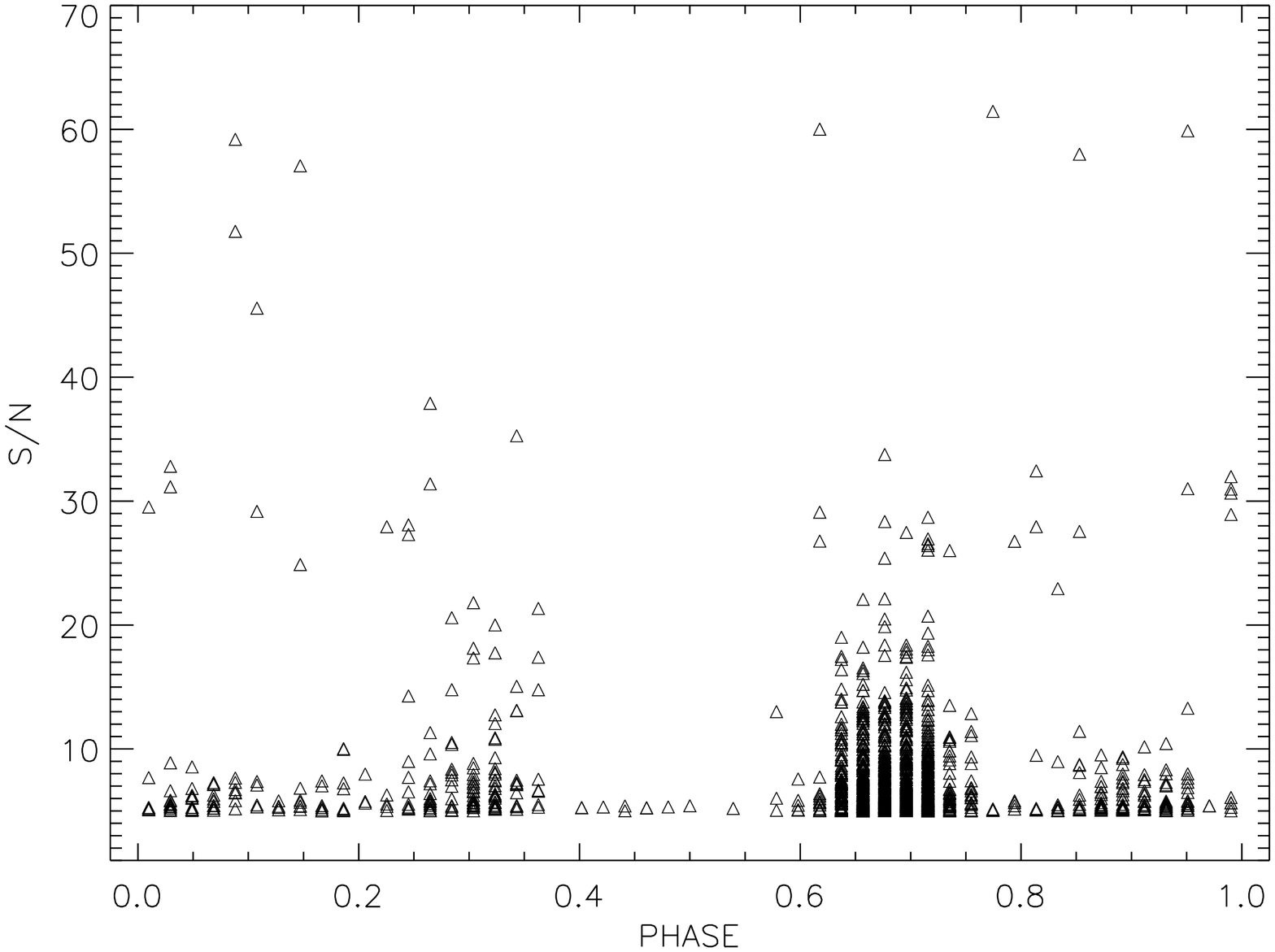}
\caption{Upper: pre-fit phases of pulse arrivals in the night of November 27-28, 2003.
Data from a single scan were taken. Phases 0 and -0.38 correspond to the location of
the IP and MP, respectively. Bottom: recorded strength and phases of 1318 GRPs.
Location of the radio components: HFC2: 0.05, precursor: 0.2, MP: 0.3, IP: 0.7, HFC1:
0.9.}
\end{center}
\end{figure}  

\newpage

\begin{figure}
\begin{center}
\includegraphics[angle = 90,scale =  0.5]{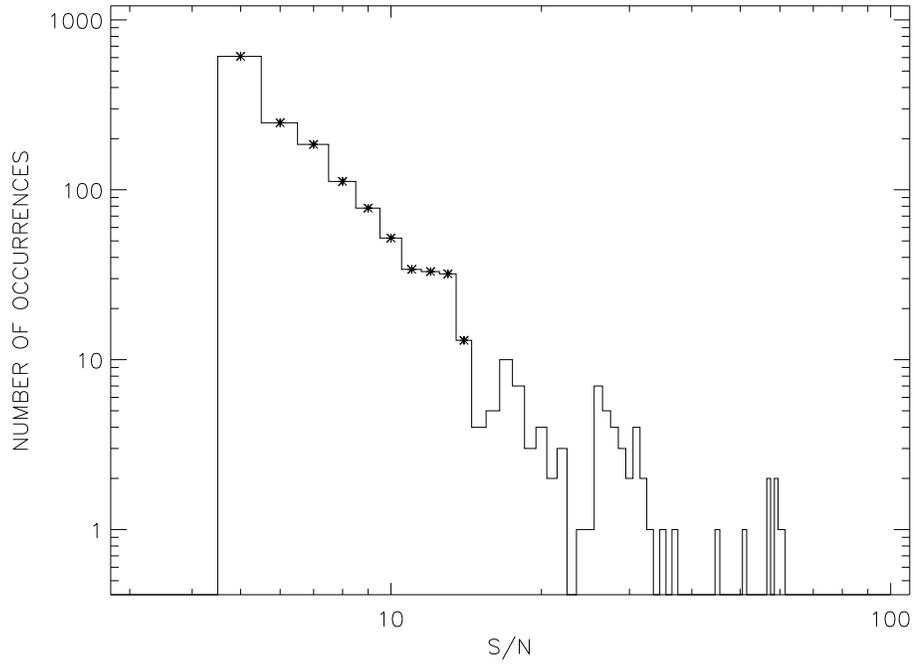}
\caption{Peak strength distribution of giant radio pulses from
the Crab pulsar regardless of their phases.  For the number of occurrences
larger or equal 10 (marked with~$\star$)
the distribution can be roughly described
by a power law $( S/N) ^{\alpha}$
with index $\alpha \sim -3.34 \pm 0.19$.}
\end{center}
\end{figure}  

\newpage

\begin{figure}
\begin{center}
\includegraphics[angle = 90,scale =  0.68]{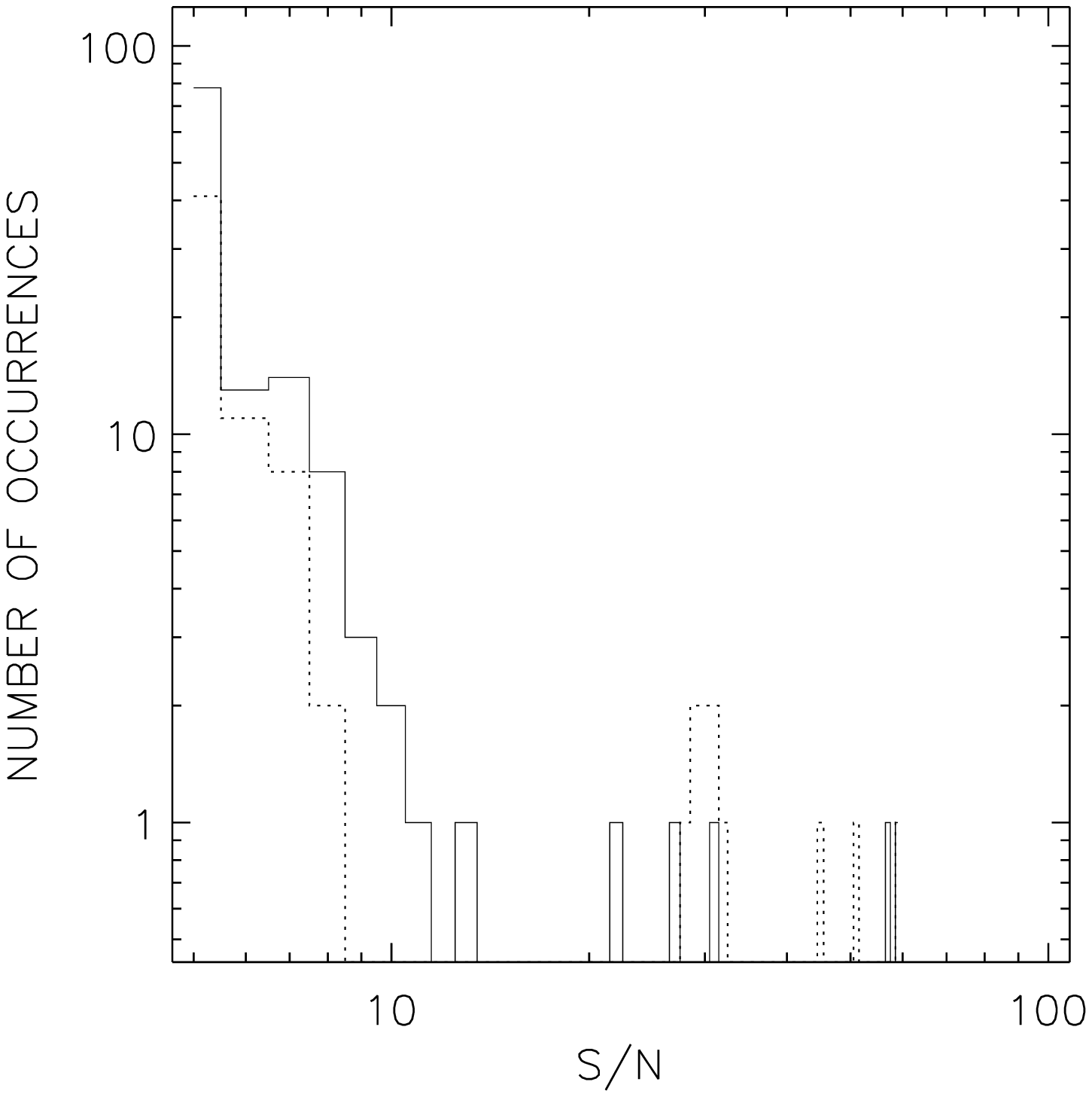}
\includegraphics[angle = 90,scale =  0.68]{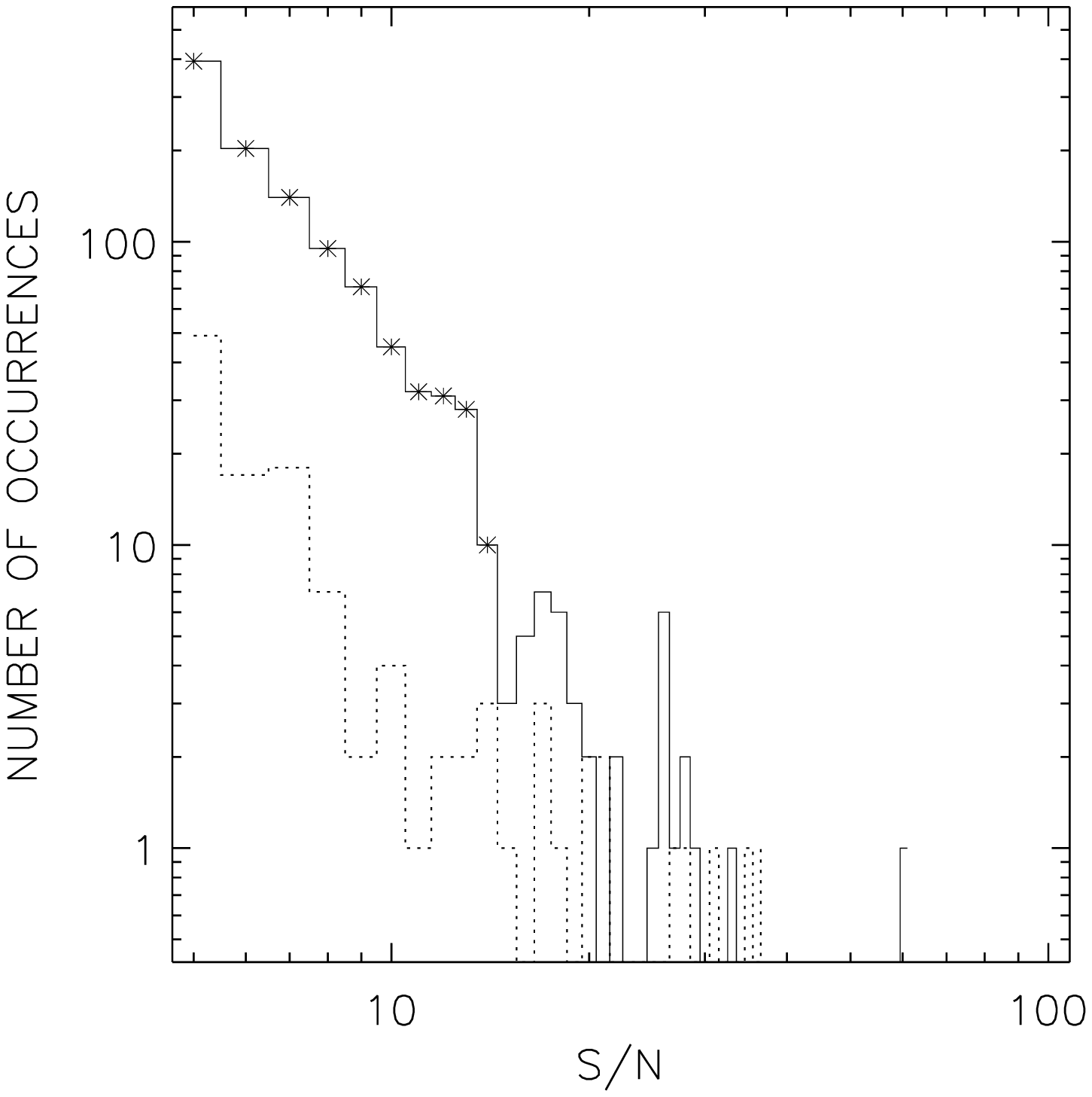}
\caption{Same as in Fig.~4, but for four phase slices, containing
HFC1 and HFC2, MP and IP, respectively. Upper panel is for
HFC1 (dotted line) and HFC2 (solid line). Bottom panel shows
MP (dotted line) and IP (solid line) components.
The giant radio pulses distribution for IP can be 
described by a power law $(S/N)^{\alpha}$ with index
$\alpha \sim -3.13 \pm 0.22$ (for the number of occurrences
larger or equal 10; marked with~$\star$).}
\end{center}
\end{figure}

\newpage

\begin{figure}
\begin{center}
\includegraphics[angle = 0,scale =  0.5]{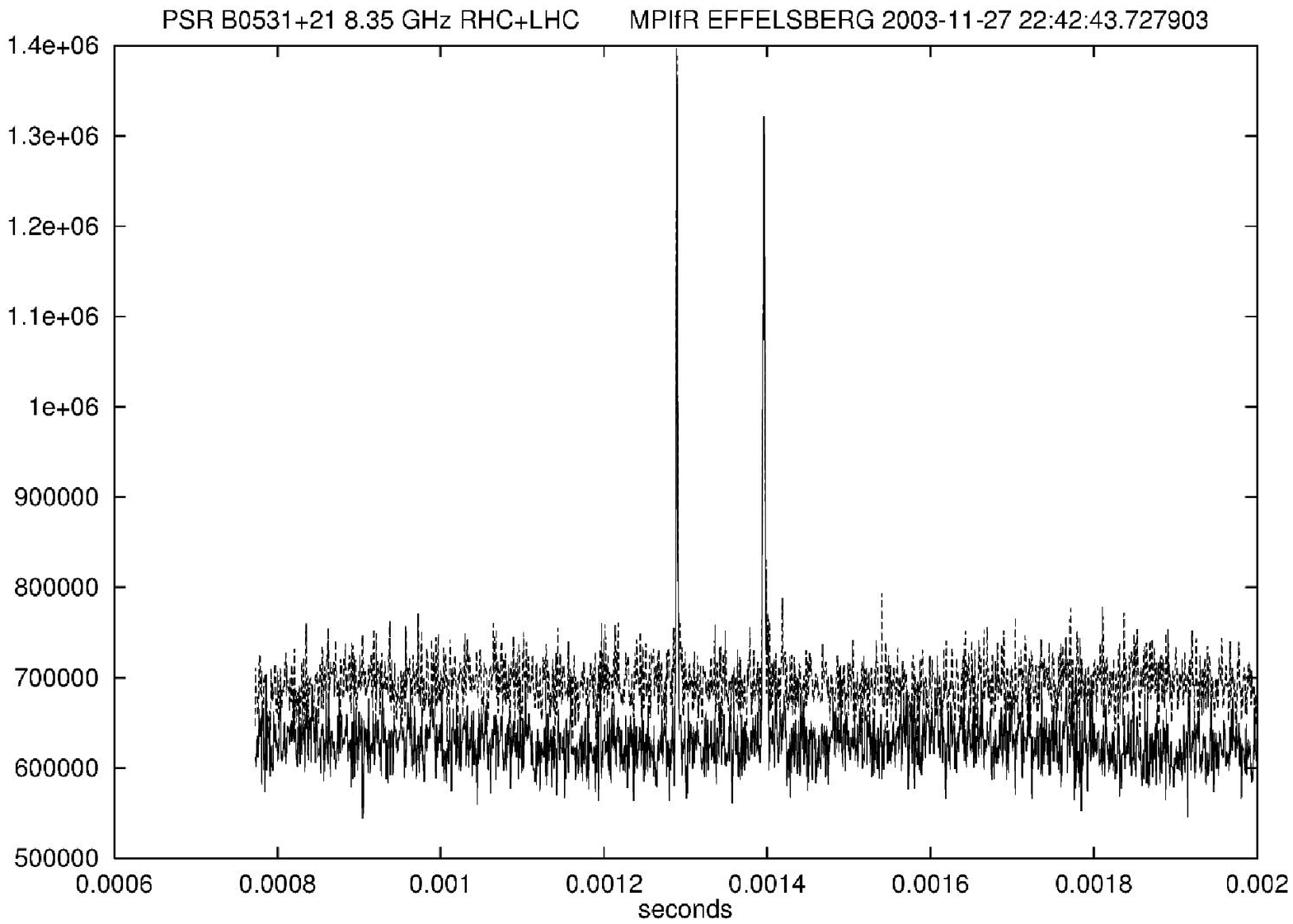}
\includegraphics[angle = 0,scale =  0.55]{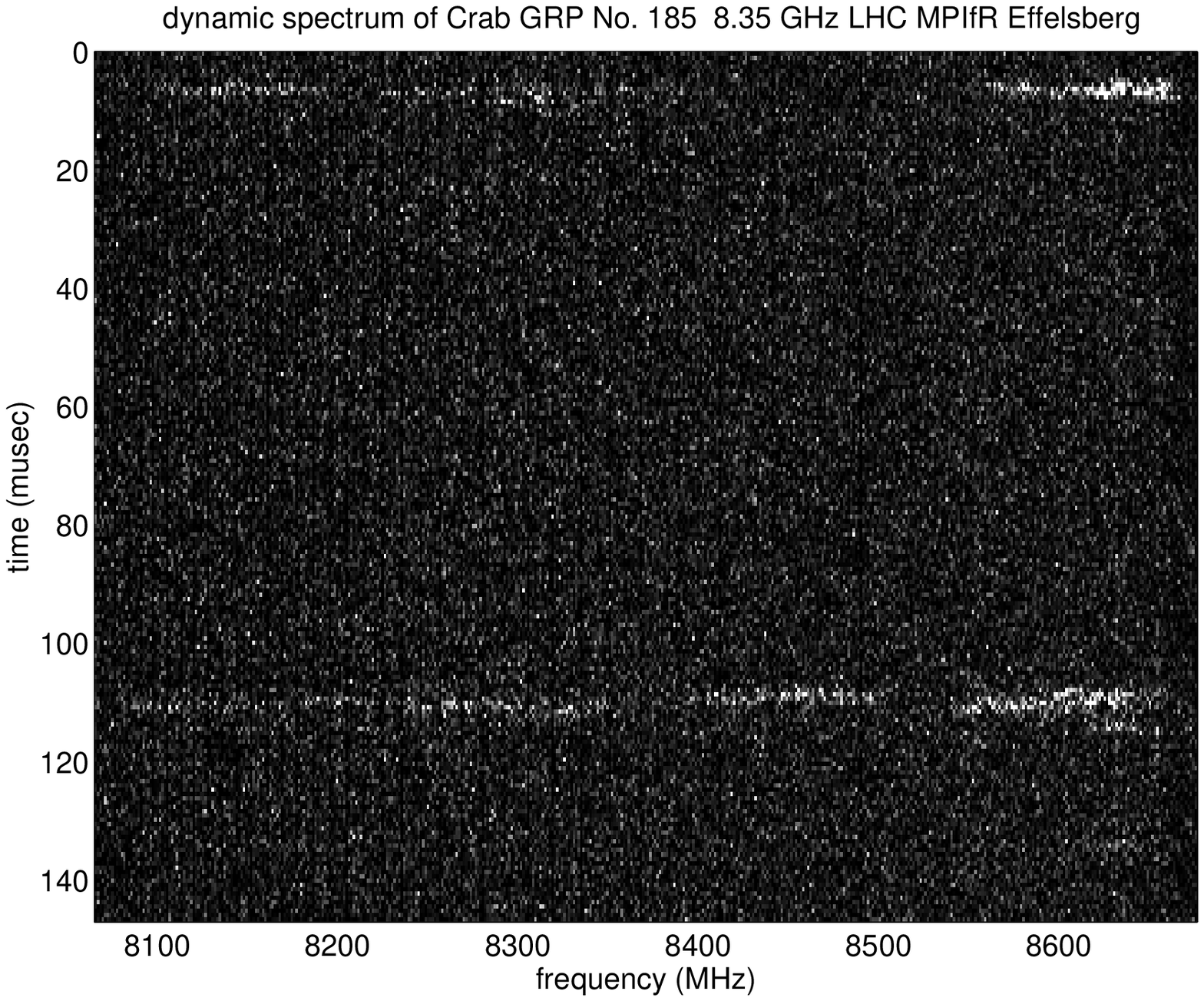}
\caption{Giant pulse with double structure at 8.35~GHz (LHC, RHC), 
peak strength $\sim$~116~Jy. Upper: de-dispersed profiles. Bottom: de-dispersed
dynamic spectrum.}
\end{center}
\end{figure}

\newpage

\begin{figure}
\begin{center}
\includegraphics[scale =  0.5]{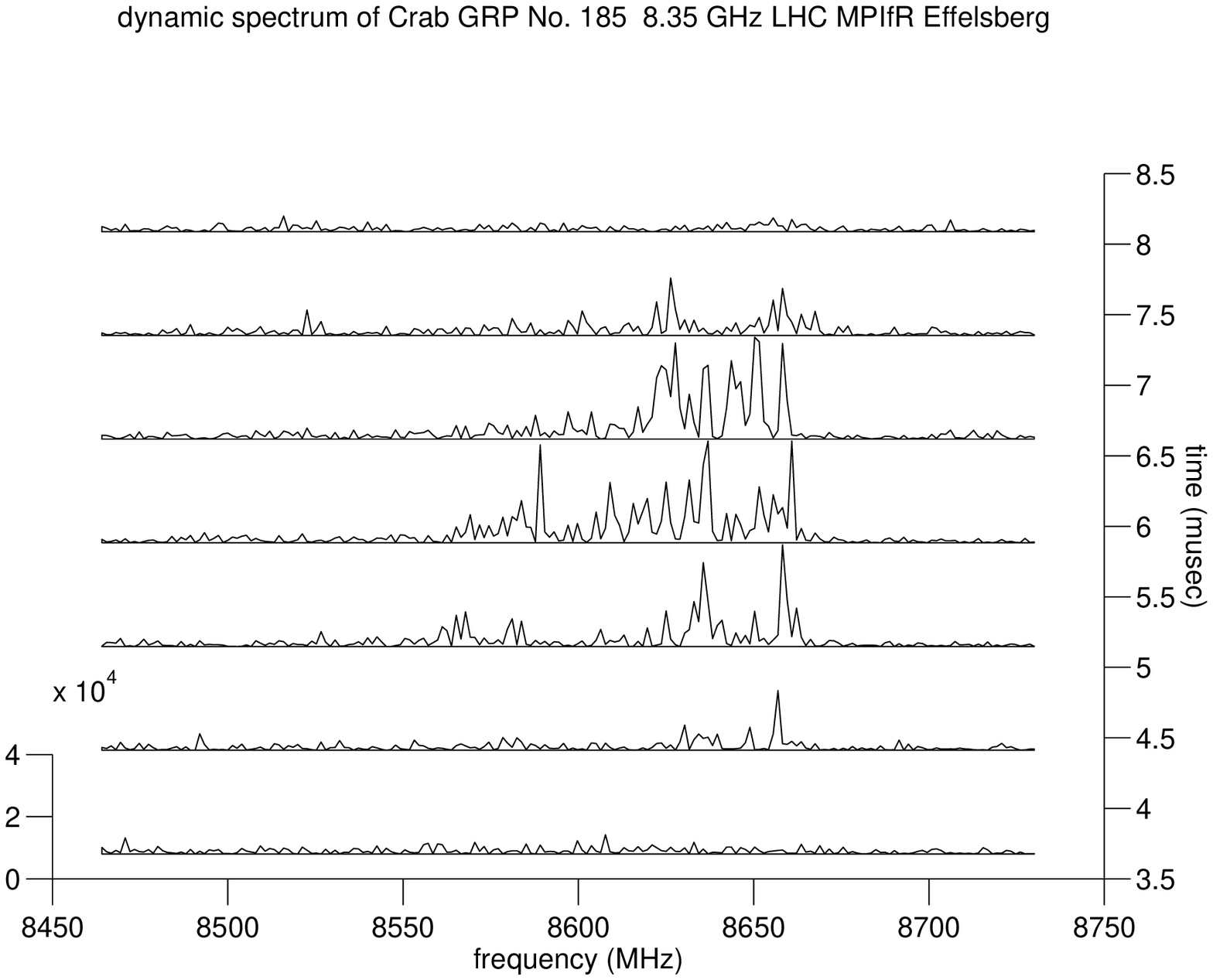}
\includegraphics[scale =  0.5]{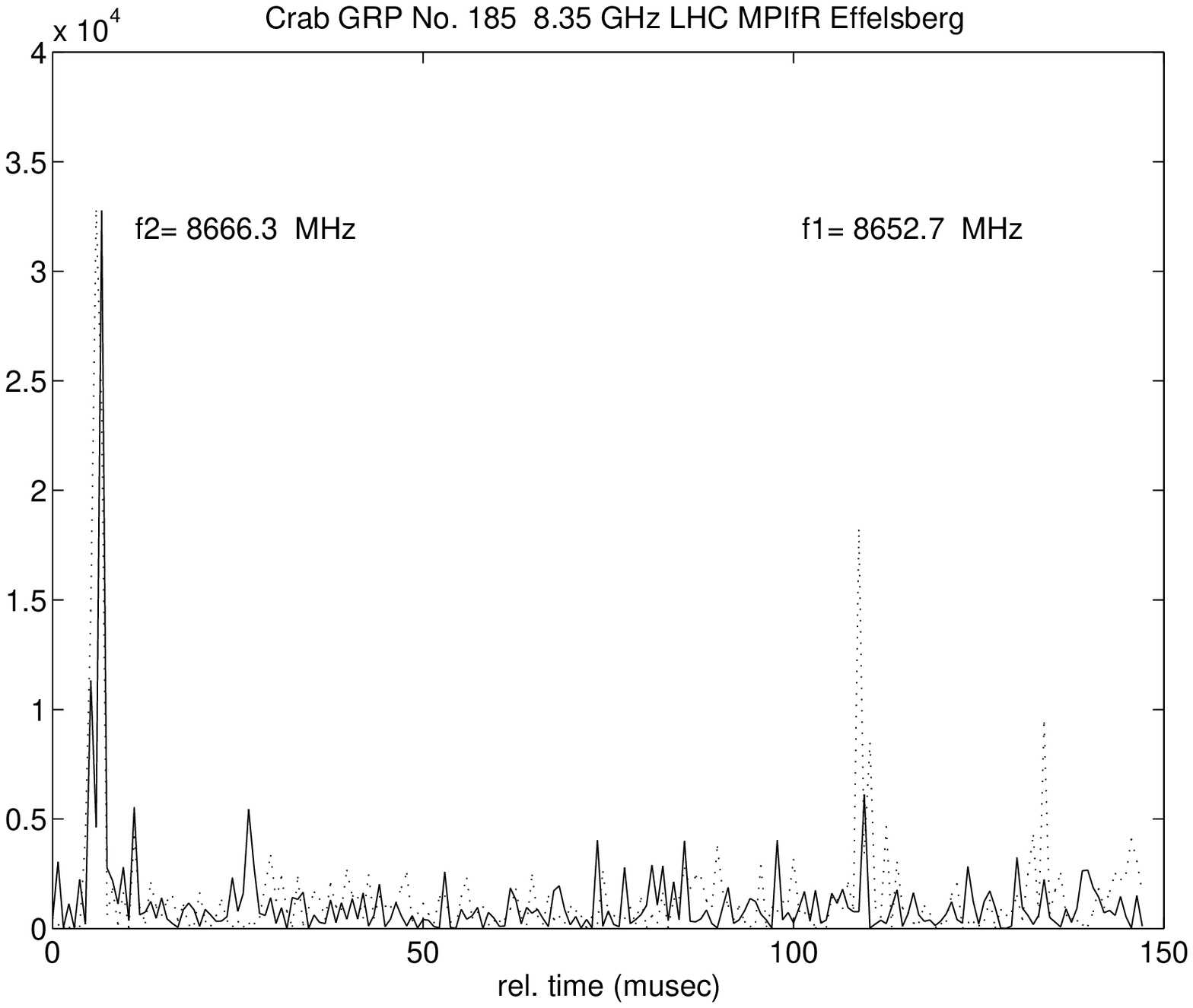}
\end{center}
\caption{Spectral structure of giant pulse components. Upper: sequence of spectra
of one GRP component. Bottom: intensity in two spectral channels separated by
13.6~MHz.}
\end{figure}

\end{document}